\documentclass[
]{ceurart}

\usepackage{enumitem}

\begin{document}

\copyrightyear{2024}
\copyrightclause{Copyright for this paper by its authors.
  Use permitted under Creative Commons License Attribution 4.0
  International (CC BY 4.0).}

\conference{Joint Proceedings of the ACM IUI Workshops 2024, March 18-21, 2024, Greenville, South Carolina, USA}

\title{Towards Full Authorship with AI: Supporting Revision with AI-Generated Views}

\author[1]{Jiho Kim}[%
orcid=0000-0002-1434-4440,
email=jihokim8@acm.org,
url=https://jihokim.dev/,
]
\cormark[1]
\address[1]{Calvin University, 3201 Burton St. SE, Grand Rapids, MI, 49546, USA}

\author[1]{Ray C. Flanagan}[%
email=rcf6@calvin.edu,
]

\author[1]{Noelle E. Haviland}[%
email=neh8@calvin.edu,
]

\author[1]{ZeAi Sun}[%
email=zs35@calvin.edu,
]

\author[1]{Souad N. Yakubu}[%
email=sny2@calvin.edu,
]

\author[1]{Edom A. Maru}[%
email=eam43@calvin.edu,
]

\author[1]{Kenneth C. Arnold}[%
orcid=0000-0003-3892-9870,
email=kcarnold@alum.mit.edu,
url=https://kenarnold.org/,
]

\cortext[1]{Corresponding author.}
%\fntext[1]{These authors contributed equally.}

\begin{abstract}
    % Large language model (LLM) based tools promise efficiency and productivity when writing. However, they can overshadow the thinking process necessary for original writing, compromising autonomy and authorship. This limitation is not intrinsic to the LLM itself but is due to the design of the chatbot user interface, which encourages reliance on an LLM for both thinking and writing. To explore alternative approaches and the potential benefits that LLMs can offer to the writing process, we prototyped a user interface that encourages writers to use the LLM to reflect on texts they wrote themselves. We conducted a pilot user study to investigate the impact of these new user interface affordances on writers' revision process. Our study results suggest that AI-generated views can help writers with discovery, audience, and clarity, but that views present interaction design challenges related to navigation, scope, and customization.
    Large language models (LLMs) are shaping a new user interface (UI) paradigm in writing tools by enabling users to generate text through prompts. This paradigm shifts some creative control from the user to the system, thereby diminishing the user's authorship and autonomy in the writing process. To restore autonomy, we introduce Textfocals, a UI prototype designed to investigate a human-centered approach that emphasizes the user's role in writing. Textfocals supports the writing process by providing LLM-generated summaries, questions, and advice (i.e., LLM views) in a sidebar of a text editor, encouraging reflection and self-driven revision in writing without direct text generation. Textfocals' UI affordances, including contextually adaptive views and scaffolding for prompt selection and customization, offer a novel way to interact with LLMs where users maintain full authorship of their writing. A formative user study with Textfocals showed promising evidence that this approach might help users develop underdeveloped ideas, cater to the rhetorical audience, and clarify their writing. However, the study also showed interaction design challenges related to document navigation and scoping, prompt engineering, and context management. Our work highlights the breadth of the design space of writing support interfaces powered by generative AI that maintain authorship integrity.
\end{abstract}


\maketitle

\section{Introduction}

Large language models (LLMs) can produce texts comparable to those written by competent human writers~\cite{brown2020LanguageModelsAre,achiam2023GPT4TechnicalReport,anil2023GeminiFamilyHighly}. Two interaction techniques currently dominate: dialogue (e.g., OpenAI's ChatGPT and Google's Gemini) and predictive text completion (e.g., GitHub Copilot). Although these paradigms work well for many tasks, in a writing context, both entail delegating some or all of the creative decision-making to the system. In other words, these systems embody the principle that \emph{the system originates the content}. For example, as demonstrated by OpenAI to introduce ChatGPT, users can specify a desired output based on a goal, such as ``help me write a short note to introduce myself to my neighbor,'' and the chatbot will respond with a generated text, such as a generic template for a note to a neighbor\footnote{https://openai.com/blog/chatgpt}. This emphasis on LLMs originating written content diminishes the user's creativity and independence in the writing process, potentially hindering the development of original ideas or unduly influencing the perspectives expressed~\cite{Arnold2018:sentiment-bias, jakesch2023CoWritingOpinionatedLanguage}.

In this work, we propose a human-centered approach to designing interactive LLM-powered systems that support the writing process. We introduce Textfocals, a UI prototype conceived with the principle that \emph{the user originates the content}. Our design prevents the LLM from modifying the user's writing and discourages situations where users might copy and paste content generated by the system into their writing. Instead, our interface motivates users to reflect on their text with LLM-generated summaries, questions, and advice on writing (which we refer to as LLM \emph{views}), helping them discover opportunities for improvement or elaboration. Specifically, compared with a general-purpose dialogue UI like ChatGPT, our prototype provides two UI affordances to reduce both the \emph{physical} and \emph{cognitive} load of generating views. First, since writers often focus their revision effort on small sections such as paragraphs, Textfocals adapts its views to the region where the writer is currently revising. This contextual adaptation, done by integrating into a professional text editing tool, aims to reduce physical effort compared to having the user provide appropriate context to a dialogue interaction (e.g., by copy and paste of users' writing into the chatbot). Second, since it is cognitively challenging to compose prompts that get LLMs to generate appropriate views, we provide scaffolding for users to select or modify pre-engineered prompts, which also helps users discover LLM capabilities relevant to their current revision needs. Previous works have explored a UI that presented LLM-generated summaries in an interactive sidebar within a text editor to assist with reverse outlining~\cite{dang2022TextGenerationSupporting}, and a UI for supporting template-based prompt engineering for better user-defined feedback on writing~\cite{benharrak2023WriterDefinedAIPersonas,macneil2023PromptMiddlewareMapping}. However, Textfocals is the first prototype to explore UI affordances for using LLMs to facilitate human reflection and discovery for making independent and self-directed revisions in writing.

We conducted a formative user study with four participants to qualitatively evaluate the effectiveness of LLM-generated summary, inquisitive, and advisory views in supporting writing revision. Our study revealed that LLM views can help users develop underdeveloped ideas, cater to their rhetorical audience, and improve clarity in writing. Specifically, participants found \emph{summary} views helpful for restructuring and expanding their writing (consistent with Dang et al.'s~\cite{dang2022TextGenerationSupporting} findings), \emph{inquisitive} views helpful for considering audience interpretations, and \emph{advisory} views helpful for discovering areas for potential improvement.

In summary, our work makes the following contributions: 

\begin{itemize}
    \item \textbf{Design and implementation of Textfocals, a UI prototype for a writing tool that facilitates reflection and discovery for making independent revisions in writing.} The prototype provides the following UI affordances: (1) a scaffolding for LLM view-generating prompts that users can use or customize and (2) a sidebar of views that contextually adapts to the part of the text being revised by the user.
    \item \textbf{Formative insights from a user study.} Our findings show that Textfocals can help users maintain creative control and authorship of their writing through LLM views, which help users refine incomplete ideas, tailor their writing to their intended audience, and improve the clarity of their writing.
\end{itemize}

\section{Background and Related Work}

\subsection{What Does It Mean to Revise?}

Our definition of revision largely aligns with that of Jill Fitzgerald, who defines it as “making any changes at any point in the writing process. It involves identifying discrepancies between intended and instantiated text, deciding what could or should be changed in the text and how to make desired changes, and operating, that is, making the desired changes. Changes may or may not affect meaning of the text, and they may be major or minor. Also, changes may be made in the writer’s mind before being instantiated in written text, at the time text is first written, and/or after text is first written”~\cite{fitzgerald1987ResearchRevisionWriting}. In other words, revision means critically examining and evaluating, which we refer to as \textit{reflection}, and identifying any opportunities for improvement or further development, which we refer to as \textit{discovery}, and then making the appropriate changes. This can occur at any stage of the writing journey. Our system leverages an LLM to support reflection and discovery to facilitate autonomous revision in writing.

\subsection{Reflection and Discovery as Facilitators of Independent Writing}

In the introduction, we defined \textit{views} as LLM-generated summaries, questions, and advice on writing. This definition is informed by research in writing, which indicates that receiving feedback from peers and teachers can improve the quality of revisions in composition~\cite{fitzgerald1987ResearchRevisionWriting,horning2006RevisionHistoryTheory,murray2003CraftRevision}. Although the terms \emph{views} and \emph{feedback} may have similar underlying semantics, we use the former term to emphasize the conclusions of the writing research conducted by Nelson and Schunn~\cite{nelson2009NatureFeedbackHow}. They found that the quality of revision depends on the type of feedback received and whether it contributes to the comprehension of the problem in the text, leading to high-quality revision. Specifically, feedback that summarizes the text helps writers see where the reader might not understand what the writer intended. Additionally, feedback that identifies an issue and offers a potential solution, especially during the earlier draft stages, helps the writer comprehend the problem in the text. Therefore, not all types of feedback are necessarily useful views, and the role of views is to help the writers scrutinize the text at hand to identify parts of the text that could be improved and understand why they could be improved.

Furthermore, previous work by Dang et al.~\cite{dang2022TextGenerationSupporting} found that automatic summaries help facilitate reflection on writing. However, their system primarily supported structural revision, while our system aims to support content revision. Moreover, a related work by Benharrak et al.~\cite{benharrak2023WriterDefinedAIPersonas} investigates a system that helps users define the persona of their rhetorical audience, which can be used to prompt GPT-3.5 to adopt a specific persona when generating feedback. They found that it motivated users to make changes to their text that would resonate better with their readers.
%However, our system, which also utilizes GPT-3.5 to generate advice and thought-provoking questions, assumes only the default persona: ``You are a helpful assistant.'' Nevertheless, 
Our system aims to generate views that could potentially facilitate thoughtful examination and analysis (i.e., reflection) and potentially lead to the discovery of new insights. This, in turn, potentially motivates high-quality and original revision that better serves the audience. Indeed, according to Hayes, ``revisions that are stimulated by the discovery of new connections, new ideas, or new arguments seem intrinsically more interesting. Such revisions are likely to be associated with improvements in the substance rather than the form of the text. They may mark those occasions when the writer learns something through the act of writing. Indeed, when I was revising this text, there were several occasions when revisions were triggered by discoveries. In many cases, these discoveries were stimulated by editor's comments but, in others, they were stimulated simply by re-reading the text''~\cite{hayes2004WhatTriggersRevision}.

\begin{figure}
  \centering
  \includegraphics[
  height=9cm,
  keepaspectratio,
  ]{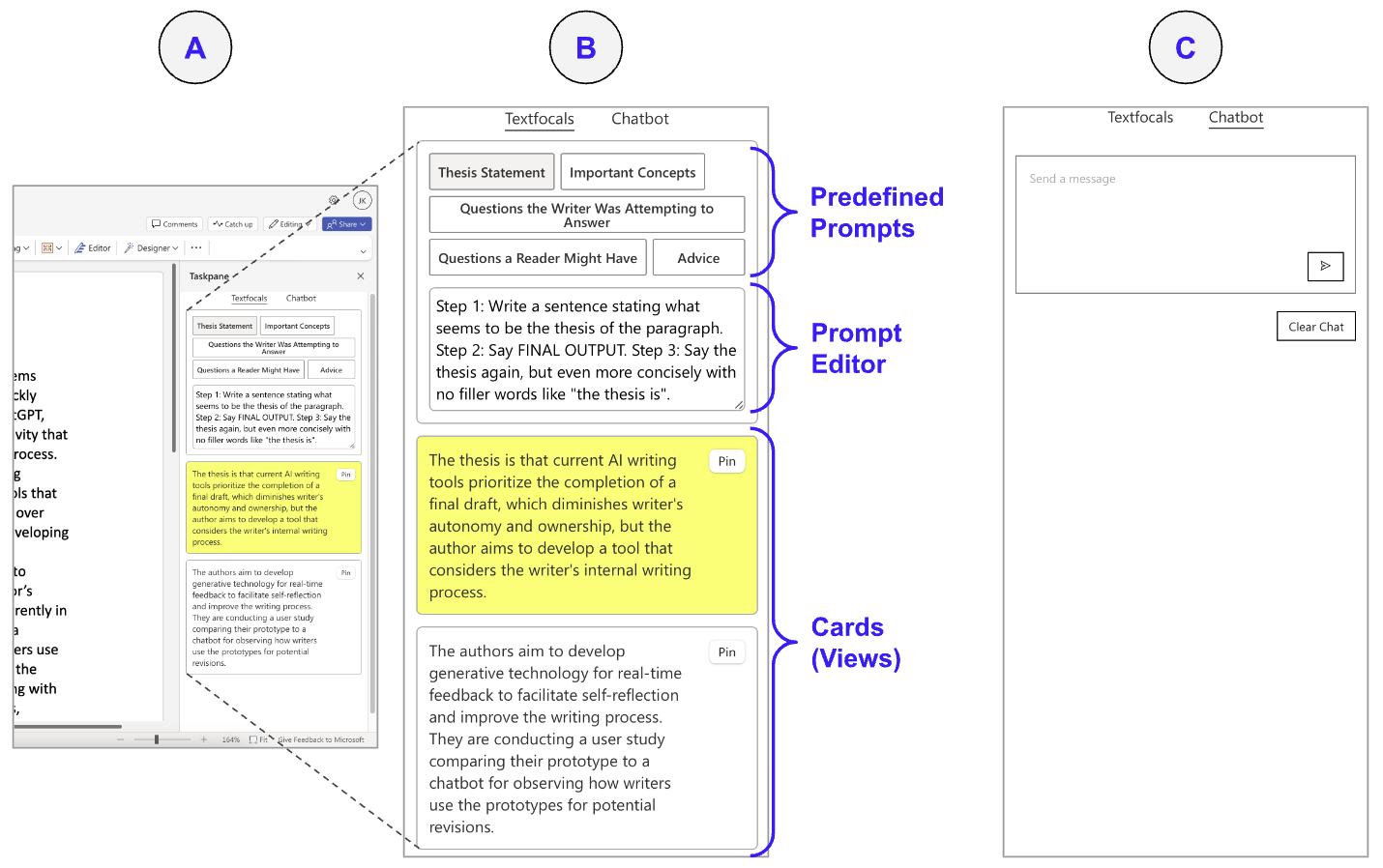}
  \caption{Our Textfocals prototype implemented as a Microsoft Word add-in. Users can interact with the prototype as a ``Taskpane'' in their Microsoft Word document, as shown in (A). Textfocals includes grouped buttons of predefined prompts, a prompt editor, and cards (views), as shown in (B). The chatbot prototype developed for the pilot user study has a conventional chatbot UI, as shown in (C).}
\end{figure}

\section{Design and Implementation}

As outlined in the previous section, we have identified reflection and discovery as core cognitive processes linked to independent revision in writing. However, even though LLM-powered writing tools such as OpenAI’s ChatGPT and Google’s Gemini can generate text similar to that of skilled human writers, these systems burden users with the responsibility of asking the right questions to the LLM for the best response (i.e., prompt engineering) and incorporating the response into their writing while still maintaining their desired level of authorship and creative control over it. To address these challenges, we aim to provide (1) a menu of preprogrammed but customizable prompts that ask the LLM to ``observe'' the text instead of producing text and (2) a sidebar of cards that help users interact with the LLM's response (i.e., views) in the context of their writing. Figure 2 illustrates the overall flow of interaction in Textfocals.

Our prototype was developed as a Microsoft Word taskpane add-in using React and Microsoft's JavaScript API. The add-in listens for cursor position changes in the document, queries the document for the text of the containing paragraph, then queries the backend (a Python FastAPI server) with the prompt and document text. The backend, in turn, queries GPT-3.5 using the OpenAI API and streams its output back to the frontend, which does lightweight parsing (such as Markdown rendering) to show the generated views.

\begin{figure}
  \centering
  \includegraphics[width=\linewidth]{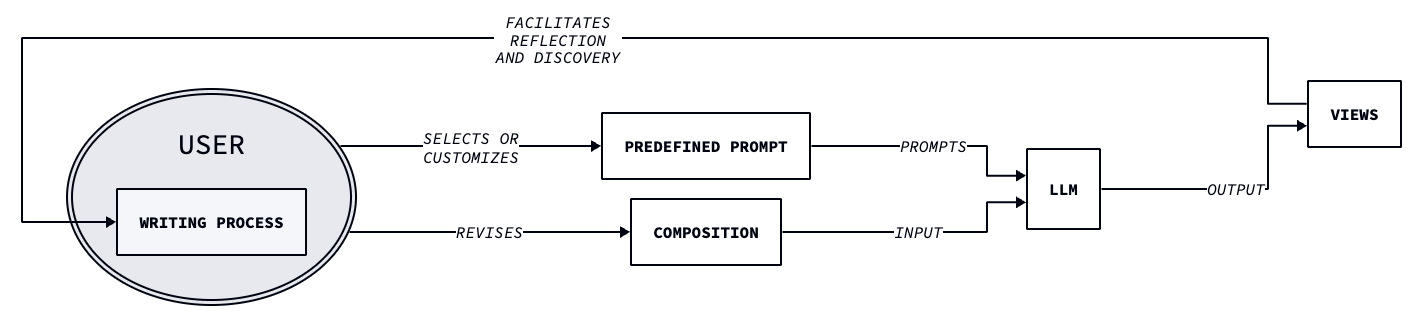}
  \caption{An overview of a user’s interaction with Textfocals: the user selects or customizes a predefined prompt. This prompts the LLM to generate views that the user can utilize for reflection and discovery in their writing process. Ultimately, this may result in the user’s revision of the composition.}
\end{figure}

\subsection{Predefined Prompts and Prompt Editor}

To support users with prompt engineering, Textfocals includes a button group to select from predefined prompts (see Figure 1). These prompts request the LLM to provide observations on the user's text (i.e., views) instead of generating document text. The prompt buttons are labeled with short summaries of their functionality, but we additionally provide a prompt editor that allows users to see and edit the underlying predefined prompts (see Figure 1). Our intention behind this approach is to inspire users to modify the prompts, encouraging them to ask the LLM to empower their writing process instead of writing for them. We have identified multiple categories of views that can be useful for reflection and discovering areas for improvement in writing. These categories include summary views that summarize the main thesis and important concepts, inquisitive views that pose questions about the text from both a reader's and writer's standpoint, and advisory views that provide suggestions for writing, both on a superficial and substantial level. For each category of views, we preprogrammed the prompts as follows:

\medskip\noindent
Summary views:\footnote{These prompts reflect a lightweight attempt at chain-of-thought prompting; we modified the system to filter out the model's internal dialogue by discarding text before ``FINAL OUTPUT.'' Additional prompt engineering could further improve these prompts.}
\begin{itemize}
    \item \textbf{Thesis Statement:} \textit{``Step 1: Write a sentence stating what seems to be the thesis of the paragraph. Step 2: Say FINAL OUTPUT. Step 3: Say the thesis again, but even more concisely with no filler words like `the thesis is.'''}
    \item \textbf{Important Concepts:} \textit{``Step 1: List 10 important concepts in this paragraph, in the format 1. Concept: [concept as a complete sentence] Relevance: [relevance score, 10 best]. Step 2: Output FINAL OUTPUT, then a new line, then a Markdown unordered list with the 3 concepts with highest relevance, in short phrases of 2 or 3 words.''}
\end{itemize}

\noindent
Inquisitive views:
\begin{itemize}
    \item \textbf{Questions the Writer Was Attempting to Answer:} \textit{``List 2 or 3 questions that the writer was attempting to answer in this paragraph.''}
    \item \textbf{Questions a Reader Might Have:} \textit{``As a reader, ask the writer 2 or 3 questions about definitions, logical connections, or some needed background information.''}
\end{itemize}

\noindent
Advisory views:
\begin{itemize}
    \item \textbf{Advice:} \textit{``What advice would you give the writer to improve this paragraph? Respond in a bulleted list.''}
\end{itemize}

\subsection{Sidebar of Cards}

To support user interaction with the generated views in the context of their writing, we implemented a scrollable sidebar consisting of interactive cards (see Figure 1). Our design draws inspiration from Dang et al.’s ``margin annotations'' design concept, which presents automated summaries as cards on a sidebar~\cite{dang2022TextGenerationSupporting}. These cards are an intermediate interface connecting the user’s writing with the LLM’s views. To initiate the interaction process, the LLM is given the first paragraph of the user’s writing as input and the predefined ``Thesis statement'' prompt discussed earlier. When a specific section of text is later focused or selected by the user in the document, the containing paragraph containing it internally snaps to the nearest paragraph and is given as input for the LLM. The LLM also generates views for the preceding and succeeding paragraphs, enabling users to browse through the views and interact with them within the context of neighboring paragraphs. Additionally, hovering over a card enables them to highlight the associated paragraph by simply allowing for easy navigation of their document based on the generated views.

%% TODO: this would be a good place to discuss "summary", "inquisitive", and "advisory
\section{Study Results}

To explore how LLM-generated views shape user's writing processes, we conducted a formative user study with N=4 participants\footnote{Our Institutional Review Board approved our study procedures.}. Participants, consisting of both staff and professors, were recruited from our university. Many of the participants identified as frequent writers who actively seek feedback on their writing. Each participant brought a draft of writing they were working on, typically about a page. Specifically, P1 was working on a newsletter, P2 was working on a grant proposal, P3 was working on an argumentative essay, and P4 was working on a blog post. They interacted with the Textfocals and chatbot interfaces (see Figure 1). Participants were encouraged to verbalize their thoughts. We analyzed transcripts and screen recordings to gain the following initial insights.

\subsection{How LLM Views Can Help Writers}

Three common themes that emerged in how our study participants used LLM-generated views were (1) discovering underdeveloped ideas, (2) catering to their audience, and (3) identifying opportunities to improve clarity.

\subsubsection{Discovering Underdeveloped Ideas}

Textfocals included prompts for summarizing the thesis and important concepts in each paragraph (as shown in Section 3.1). Some of our participants found that these summaries could not only help restructure the document (as Dang et al.~\cite{dang2022TextGenerationSupporting} found) but also help draw attention to parts of their writing that they had not previously considered, thereby helping them identify ideas that could be further developed. For example, when P4 requested a list of important concepts in a short paragraph, the system identified one (``tagging the creator'') that they had mentioned in passing but realized that “Maybe that’s an idea that I could develop a little bit more,” continuing, “I’m really pleasantly surprised that… it could actually help [me] generate new ideas, not just by asking it to give me ideas, but by engaging with it....” This shows that summary views could help users in expanding their writing by highlighting specific areas that could potentially be explored in more depth.

\subsubsection{Catering to the Rhetorical Audience}

Summary views also helped the participants compare their readers’ understanding of their writing with their intended message and identify any discrepancies. The participants treated the summary views as though they had been written by an external reader. For example, P4 commented, “I think one thing that’s helpful [about the view] is it helps me understand how [my audience] might get one level of understanding of what they’re reading here\ldots{} that AI can help me understand how readers might take in my writing and misunderstand it, or understand different levels of it.” This shows that summary views could help users revise their composition so that they clearly convey their intended message to their intended audience.

Textfocals also included inquisitive views that ask personified questions about the participant’s writing. These views also helped users acquire a deeper understanding of how their intended audience might interpret their writing differently. For instance, P3 stated that they would make changes to their composition if their current writing could not adequately answer the questions posed by the view, which could reasonably be asked by their audience. P3 likened their experience of reflecting on their writing using the view to the following analogy: “You can sometimes have a vision of ‘I’m writing this piece about this,’ and then someone else would be like, ‘Well, it’s actually about this,’ and you go, ‘Oh, yeah, it is.’” This shows that inquisitive views could also help users to revise their document to better serve their rhetorical audience.

\subsubsection{Identifying Opportunities to Improve Clarity}

Although the participants found advisory views generally helpful in identifying both superficial and substantive improvements that could be made to their writing, many participants also expressed a desire to see a practical example demonstrating how these improvements could be implemented in their writing. For example, after reading an advisory view that showed “Reorganize the paragraph to flow logically and smoothly,” and reflecting on the associated text, P4 thought out loud and said, “How can I be more logical?... I’m curious to know what it means by ‘logical’,” implying that they would have appreciated an example. Such demand also led to an unexpected interaction. Most notably, while reflecting on their writing using the view: “Break the paragraph into smaller, more concise sentences to improve readability and flow,” P2 asked, “How do I ask it [Textfocals] to do that? I want the program to show me what it suggests I do.”  Thus, for clarity revisions and other advice, system responses should include examples.

Dialogue-style interaction may provide a natural way for users to ask for these sorts of examples, as long as the dialogue agent is given access to sufficient context from the document. For example, P3 copied and pasted text from AI views into our chatbot interface multiple times during the study to get a practical example of the suggestion shown in the view. An improved design might provide an affordance to begin a conversation about any view that is shown, or about any part of the document. 

\subsection{Interaction Design Considerations}

Our pilot study revealed several design challenges and opportunities for how systems can present views and allow users to interact with them.

\subsubsection{Display and Navigation of Views}

When systems can provide feedback that requires significant space to show, and can apply to areas of different scales in the user’s document, designers need to consider how the relationship between document and feedback is visualized and what affordances the system provides for changing what area of the document they are seeing feedback on. In our study, participants requested views focused on various parts of their document, but not all of the feedback could fit on their screens at once.

Our prototype used highlighting to visualize the part of the document that each view was associated with, but participants found this confusing because the highlights were not directly visually linked with the views and because highlighting in other situations often indicates relevance rather than focus. The cursor interaction was a confusing and misunderstood feature for the participants. For example, P1 misinterpreted the yellow highlight of the text when they hovered the cursor over it as an indication of an error and said, “I’m not quite sure if it’s highlighting it because it’s good or bad.” Similarly, P3 misunderstood the yellow highlight of the card as an indication of the most relevant view, and commented, “If I look at those and especially the ones that are marked in yellow, that I’m assuming are the highest relevancy.” Other participants, P2 and P4, commented that the feature was confusing and unintuitive, and it took some time for them to get used to the feature. So, in situations where the sidebar needs to indicate what part of the document is associated with a sidebar element, a subtle and achromatic outline might be better so as not to imply additional meaning.

\subsubsection{Scoping Views to Parts of the Document}

When using the sidebar, some participants found it unclear which selection of the text (i.e., the scope of the text) was being provided as input to the LLM. For instance, while scrolling through the view cards, P1 asked whether the prototype was “looking at” the entire document or only the part they had selected. Likewise, P2 asked if the prototype was “looking at” the entire document or only the paragraph they had selected. This suggests that users would have found it useful to have an explicit visual cue that made it clear which subset of the text the LLM was using as an input to generate the views. 

% Hence, we suggest incorporating a text navigation feature in the prototype by adding on-screen directional controls with labels. This could help facilitate navigation between different text scopes (e.g., paragraphs or sentences) and also provide a clear indication of the input text scope.

\subsubsection{Prompt Flexibility}

All of our participants used several different predefined prompts for their views, demonstrating the usefulness of variety in views. Some participants additionally tried editing the predefined prompts or writing their own to address issues or needs that go beyond the predefined prompts. For example, one participant edited the ``Questions a Reader Might Have'' prompt to specify the type of reader; another tried repurposing the view functionality to request that the LLM improve the text for them. However, composing and managing these prompts was challenging for users. A fill-in-the-blanks approach such as that taken by FeedbackBuffet~\cite{macneil2023PromptMiddlewareMapping} could address part of this challenge, but users may also need support to craft novel types of prompts and recall them later. Further research on helping people express and refine their goals to AI systems could refine this interface.

\subsubsection{Contextual Information}

Our participants sometimes found the LLM-generated views irrelevant because the view did not account for the information presented elsewhere in the document. For example, the LLM would sometimes generate questions that immediately adjacent paragraphs already addressed. This behavior may have emerged in our prototype because we used only the context of the current paragraphs for generating views (since including the entire document might have exceeded the context length limit of the models we were using). Future work should consider including more context, whether by using LLMs with longer context lengths or by adaptively summarizing the context as needed. Contextual information could also include specification of the audience and other aspects of the rhetorical situation. For example, P2 complained that one of the inquisitive prompts included a question that the audience of the text would already know the answer to. Future work could explore allowing writers to specify this information, perhaps in the form of personas~\cite{benharrak2023WriterDefinedAIPersonas}.

Our participants also expected that chat conversations in the sidebar of a document should have the content (at least the currently visible part) available for context. Our chatbot prototype did not include this, but future studies should correct this oversight.

%\subsection{Lack of Document Context Integration}

%Insufficient attention was given to the design of the chatbot, resulting in an incomplete prototype. One example of this was the oversight of including the user’s document as a preloaded context for the LLM that powers the chatbot. As a result, the chatbot’s ability to generate relevant and contextual output was limited. Interestingly, P4 chose to copy and paste their entire document and send it as a message to the chatbot without being informed of its benefits. They mentioned that they did this to “ask questions” about the document. Unknown to them, this action helped the participant to receive more relevant and contextual responses from the chatbot. Therefore, we suggest improving the prototype so that the document is incorporated as context before the user initiates an interaction.

\section{Discussion}

Revision is a writing process that involves the critical examination and evaluation of the writing (i.e., reflection) to identify opportunities for improvement or further development (i.e., discovery) and make the appropriate changes~\cite{fitzgerald1987ResearchRevisionWriting}. By presenting the predefined prompts, which were crafted to prompt the LLM to generate what it “observes” instead of what it can replace or continue, it enabled the users to consider the LLM’s output as an external perspective that they could use to reflect on their writing and make discoveries. We found preliminary evidence that the participants in the pilot user study found the output generated in this manner useful. It helped them draw attention to concepts in their writing that they could develop further. It also assisted them in identifying gaps between their intended message in the writing and their audience’s understanding, allowing them to gain insights on how to cater their writing to their intended readers. Even though these observations are not a comprehensive list of how one can reflect on their writing and make discoveries, we note that they enable users to identify what Hayes refers to as “new connections, new ideas, or new arguments\dots{}” which propels “improvements in the substance rather than the form of the text”~\cite{hayes2004WhatTriggersRevision}. Therefore, our results, at least preliminarily, show that Textfocals can facilitate reflection and discovery that supports substantial revision, and help users write for their audience.

Furthermore, our results also underscore the effectiveness of predefined prompts and the prompt editor in guiding participants to prompt the LLM to generate external viewpoints on their writing rather than replacing or continuing their writing. This UI affordance effectively enables participants to leverage the LLM to empower their thinking process in writing rather than relying on the LLM to replace their own thoughts and ideas. Indeed, Hayes also notes that substantial revisions are also “associated with interesting changes\dots{} in the thinking of the text’s author”~\cite{hayes2004WhatTriggersRevision}. Therefore, our results also preliminarily demonstrate that Textfocals can help users maintain their full authorship and autonomy when revising their composition.

\subsection{Limitations}

Although our main focus was to assess Textfocals, we structured our pilot study as a within-subjects design to compare the effectiveness of new UI affordances against a chatbot interface. However, due to inconsistent study procedures, we could not draw robust conclusions. For example, even though counterbalancing was attempted to eliminate the learning effect, many participants perceived the two systems as complementary or similar, limiting our ability to assess the effectiveness of the new UI affordances.

\begin{acknowledgments}
    We would like to thank Saron Melesse for her contribution to the research. We would also like to thank Dr. Keith Vander Linden and Dr. Kristine Johnson for their feedback and advice. Additionally, we would like to express our gratitude to the participants of our pilot study. This research is supported by NSF CRII award 2246145 and the Wierenga Family Foundation Summer Research Fellowship for Sciences.
\end{acknowledgments}

%%
%% Define the bibliography file to be used
\bibliography{arnold-summer-research}

\end{document}